\documentclass[a4paper,12pt]{article}
\usepackage{amsmath}
\usepackage{amsfonts}
\usepackage{amssymb}
\usepackage{latexsym}
\usepackage{epsfig}
\usepackage{graphicx}
\usepackage{oldgerm}
\usepackage{theorem}
\usepackage{authblk}

\theorembodyfont{\itshape}

\newcommand{\SSC}[1]{\section{#1}\setcounter{equation}{0}}

\begin{document}

\title{\bf Self-Elongation with Sequential Folding \\
of a Filament of Bacterial Cells}

\author{Ryojiro Honda
\thanks{E-mail:rhonda@phys.chuo-u.ac.jp
}, 
Jun-ichi Wakita
\thanks{E-mail:wakita@phys.chuo-u.ac.jp
}, 
and Makoto Katori
\thanks{E-mail:katori@phys.chuo-u.ac.jp
}
}
\affil{\it Department of Physics, Faculty of Science and Engineering, \\
\it Chuo University, Bunkyo, Tokyo 112-8551, Japan}
%%%%%%%%%%%%%%%%%%%%%%%%%%%%%%%%%%%%%%%%%%%%
\date{21 September 2015}
%%%%%%%%%%%%%%%%%%%%%%%%%%%%%%%%%%%%%%%%%%%
\pagestyle{plain}
\maketitle
\begin{abstract}
Under hard-agar and nutrient-rich conditions,
a cell of {\it Bacillus subtilis} grows
as a single filament owing to the  failure of cell separation after
each growth and division cycle.
The self-elongating filament of cells shows
sequential folding processes, and multifold
structures extend over an agar plate.
We report that the growth process from the exponential phase
to the stationary phase 
is well described by the time evolution of 
fractal dimensions of the filament configuration.
We propose a method of characterizing filament configurations
using a set of lengths of multifold parts of a filament.
Systems of differential equations are introduced
to describe the folding processes that create multifold structures
in the early stage of the growth process.
We show that the fitting of experimental data
to the solutions of equations is excellent, 
and the parameters involved in our model systems
are determined.
\end{abstract}
                             
%\noindent KEYWORDS:
%a filament of bacterial cells,
%self-elongation with sequential folding,
%creation of multifold structures,
%crossover from one-dimensional to two-dimensional structures,
%analysis by systems of differential equations

%%%%%%%%%%%%%%%%%%%%%%%%%%%%%%%%%%%%%%%%%%%%%%%%%%%%%%%%%%
%%%  SEC1   %%%%%%%%%%%%%%%%%%%%%%%%%%%%%%%%%%%%%%%%%%%%%%
%%%%%%%%%%%%%%%%%%%%%%%%%%%%%%%%%%%%%%%%%%%%%%%%%%%%%%%%%%
\SSC{Introduction}
\label{sec:introduction}
%%%%%%%%%%%%%%%%%%%%%%%%%%%%%%%%%%%%%%%%%%%%%%%%%%%%%%%%%%

Patterns observed in bacterial colonies
growing on surfaces of semisolid agar plates
realize a variety of fractal and self-affine structures
studied in statistical mechanics and fractal physics \cite{MHKOYM04}.
In a series of experimental studies
\cite{RMWSES96,WIMM97,IWMM99,HWKYMM05}, 
it has been clarified
that the morphology of growing bacterial colonies at the macroscopic scale 
does not depend on the biological details of individual organisms, 
but depends only on environmental conditions controlled
by the agar concentration $C_{\rm a}$ and 
the nutrient concentration $C_{\rm n}$.
Cell motility is changed by varying $C_{\rm a}$ and
the growth rate is controlled by varying $C_{\rm n}$,
and then different patterns appear in different regions
in a morphology diagram drawn on the
$C_{\rm a}$-$C_{\rm n}$ plane.
On this plane, a diffusion-limited aggregation (DLA)-like pattern \cite{WS81,Mea86},
an Eden-like pattern \cite{Ede61,FV85}, a concentric-ring pattern,
a homogeneously spreading disk like pattern, 
and a dense branching morphology (DBM) pattern \cite{MHKOYM04}
have been recorded.
Since some of these patterns are observed not only in
biological systems, 
but also in chemical and physical systems, such as those in
crystal growth, aggregation processes, and viscous fingering \cite{Vic92}, 
the mechanism of pattern formation could be 
common in organisms and inorganic substances,
and a theoretical study by mathematical modeling and analysis
will be useful for understanding the underlying universal principles
\cite{VCBCS95,VZ12,BPS11,PSB13,BPS14}.

In a recent paper \cite{WTYKY15}, we have reported
the physical aspects of the collective motion
of bacterial cells observed in shallow circular pools
prepared on the surface of an agar plate.
The diameters of the pools are arranged so as not to be much larger than
the length of the bacterial cells swimming in the pools.
We used {\it Bacillus} ({\it B.}) {\it subtilis} and found six different types of collective motion, 
including one-way and two-way rotational motion along the brim of a circular pool
and collective oscillatory motion in the entire pool.
Analyzing experimental observations in 117 circular pools,
we found that these six types of collective motion
can be classified using only two parameters:
the reduced cell length $\lambda$, which is defined as the
ratio of the average cell length in a pool to the pool diameter, 
and the cell density $\rho$ in the pool. 
We obtained a phase diagram for 
the collective motion drawn on the 
$\lambda$-$\rho$ plane and predicted that
simple modeling with the two control parameters
will be able to explain the variety of collective motion
of bacterial cells.

The above results imply that physical considerations
are applicable and useful to explain the changes in the
morphology of bacterial colonies at the macroscopic scale
as well as the dynamical transitions of the collective motion
of bacterial cells at the microscopic scale
caused by environmental variations.
On the basis of them, in this paper, we will report
the experimental results and numerical analyses 
of growth processes starting from a single bacterial cell
observed in formation
of colony patterns. We tried to analyze 
the observed growth process by fractal analysis
and by using systems of differential equations.

Throughout the experiment reported in this paper, 
we used {\it B. subtilis} wild-type strain OG-01. 
Cells of this strain 
are rod-shaped (0.5--1.0 $\mu$m in diameter, 2--5 $\mu$m in length) 
with peritrichous flagella. They swim in a straight line in water 
by bundling and rotating the flagella. Under unfavorable environmental 
conditions, such as on a nutrient-poor medium or a dry agar plate, they 
become spores. When a small number of cells are inoculated on 
the surface of a semisolid agar plate, they go through a resting period
of about 7 h before starting two-dimensional colony expansion.

%%%%%%%%%%%%%%% Figure  %%%%%%%%%%%%%%%%%%%%%%%%%%%%%%%%%%%%%%%%%
\begin{figure}
\hskip 2.5cm
\includegraphics[width=0.7\linewidth]{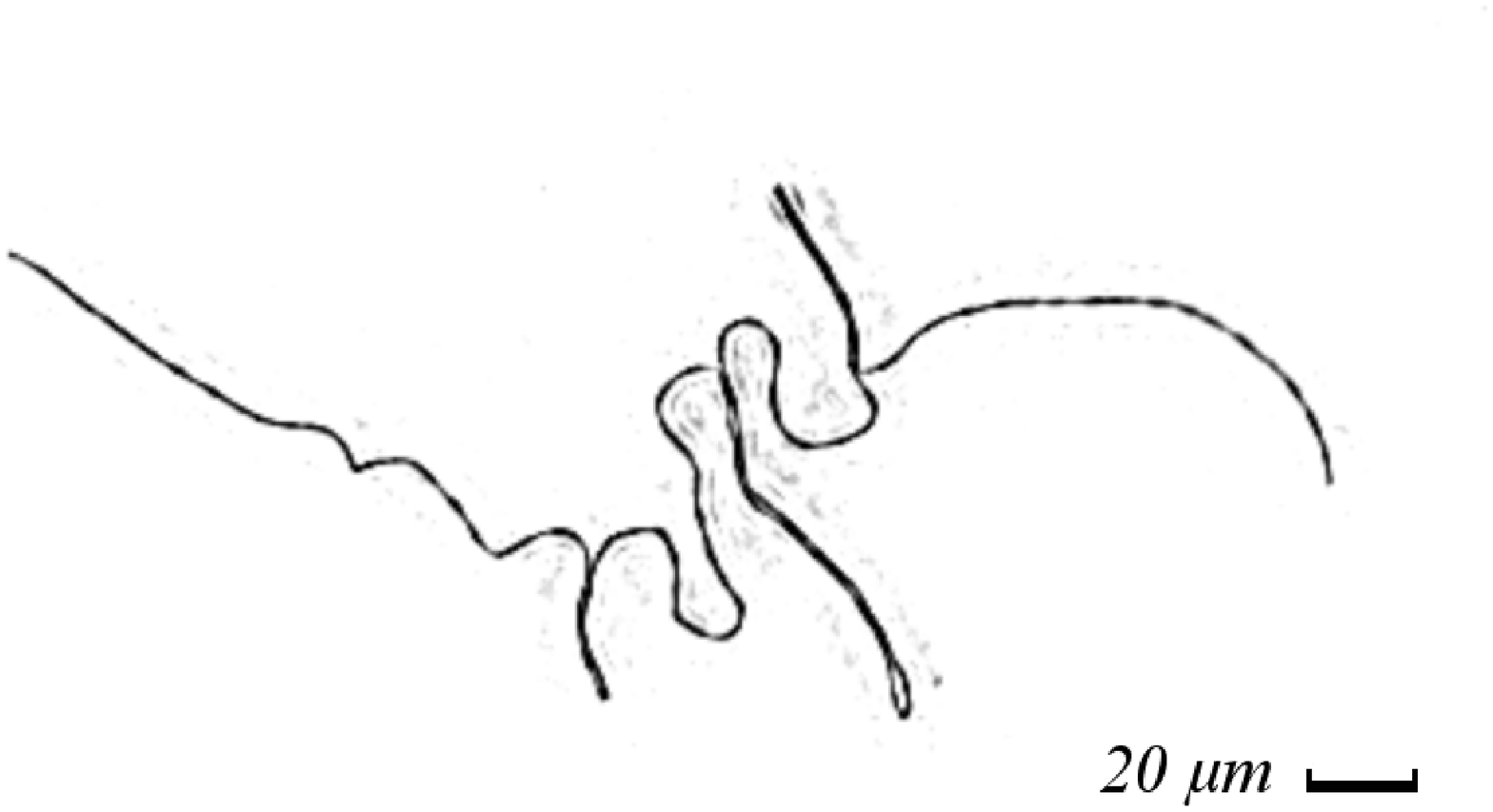}
\caption{Snapshot of the filament configuration of 
bacterial cells on an agar plate
at time $t=40$ min after the first twofold part appeared.
The scale bar indicates 20 $\mu$m.
}
\label{fig:picture1}
\end{figure}
%%%%%%%%%%%%%%%%%%%%%%%%%%%%%%%%%%%%%%%%%%%%%%%%%%%%%%%%%%

As briefly reported in previous papers \cite{WIMM97,WKMM10}, 
we observed string like objects in microscopy observations
of an inoculation spot of a bacterial suspension in the later stage of the resting phase.
In this study, 
we focus on the growth process that started from a single
bacterial cell in the resting period under hard-agar and nutrient-rich conditions.
In this case, cell multiplications are repeated with a constant
cell cycle (doubling time), but daughter cells fail to separate after each cell cycle,
although the cytoplasm has been compartmentalized by septum formation.
Then, a long filament is produced, which consists of a chain of cells
linked end to end \cite{Men76}.
Such a filament writhes as it elongates on the agar plate,
and eventually some segment starts folding and a twofold part
of the filament is created. 
Figure \ref{fig:picture1} shows a typical configuration of a filament 
in which twofold parts have been created.
Sooner or later, we will see the appearance of threefold parts, fourfold parts,
and so forth, and the filament configuration of linked cells 
becomes complicated on the agar plate.

Mendelson and coworkers have intensively studied the 
supercoiling processes performed by such cell filaments
of {\it B. subtilis} \cite{Men76,Men78, MTKL95,MSL97,Men99}. 
They are interested in the situation wherein the
filaments twist to make a double-stranded helix.
The double-stranded structure itself twists while writhing, 
eventually comes in contact with itself, and forms a supercoil.
By the repetition of such supercoiling processes,
macroscopic structures of millimeter length
are created, which are called bacterial macrofibers.
The interesting motion of macrofibers was reported
by Mendelson and coworkers. \cite{MSWG00,MST01,MMT02,MSRCT03}.
Kumada {\it et al.} also observed the growth of 
filaments of cells without separation for {\it B. subtilis},
in which the dependence of the morphology on $C_{\rm a}$ was
systematically studied \cite{KIT96}.
Such growing filamentous cells were also observed
for {\it Escherichia} ({\it E.}) {\it coli} \cite{TDWW05}.

In this paper, we study the simplest situation
wherein a filament of bacterial cells does not twist,
and hence helical structures are not formulated.
In our case, a single self-elongating filament repeatedly folds
upon itself and shows a crossover from a one-dimensional
structure to a two-dimensional structure.
In Sect. \ref{sec:experimental},
we explain the experimental setup and procedure 
for recording microscopic snapshots of cell filament configurations
for about 6 h.
We performed fractal analysis of filament configurations 
by the box-counting method.
In Sect. \ref{sec:crossover}, we report that the results
imply fractal structures in the filament configurations,
and the evaluated fractal dimension $D$ shows a crossover from 
$D=1$ to $2$ as the growth process goes from the
exponential phase to the stationary phase. 
Detailed study of the time evolution of the filament configuration
of bacterial cells is discussed in Sects. \ref{sec:measurement} 
and \ref{sec:analysis}.
In Sect. \ref{sec:measurement},
we propose a method of characterizing filament
configurations on an agar plate by a set of lengths
of the simple part and the $k$-fold parts with $k=2,3, 4, \dots$,
and the results of experimental measurements of 
these lengths are shown.
To analyze the data given in Sect. \ref{sec:measurement},
we introduce systems of differential equations
in Sect. \ref{sec:analysis} and the nonlinear fitting of data to 
their solutions is performed.
Section \ref{sec:remarks} is devoted to concluding remarks.

%%%  SEC2   %%%%%%%%%%%%%%%%%%%%%%%%%%%%%%%%%%%%%%%%%%%%%%
%%%%%%%%%%%%%%%%%%%%%%%%%%%%%%%%%%%%%%%%%%%%%%%%%%%%%%%%%%
\SSC{Experimental Procedures}
\label{sec:experimental}
%%%%%%%%%%%%%%%%%%%%%%%%%%%%%%%%%%%%%%%%%%%%%%%%%%%%%%%%%%

We observe a multiple-fission process from a single bacterial cell in the resting period 
under hard-agar and nutrient-rich conditions. 
The experimental setup and procedure are as follows.

A solution containing 5 g of sodium chloride (NaCl), 5 g of dipotassium 
hydrogen-phosphate (K$_{2}$HPO$_{4}$) and 10 g of Bacto-Peptone 
(Becton, Dickinson and Company, Franklin Lakes, NJ, USA) in 1 L 
of distilled water is prepared. The environmental parameter $C_{\rm n}$
is given as the concentration of Bacto-Peptone; $C_{\rm n} = 10$ g$\cdot$L$^{-1}$. 
Then, the solution is adjusted to pH 7.1 by adding 6 N hydrochloric 
acid (HCl). Moreover, the solution is mixed with 10 g of Bacto-Agar 
(Becton, Dickinson and Company), which determines the softness 
of a semisolid agar plate. The environmental parameter $C_{\rm a}$ is given 
as the concentration of Bacto-Agar; $C_{\rm a} = 10$ g$\cdot$L$^{-1}$. The environmental 
condition realized by these values of $C_{\rm a}$ and $C_{\rm n}$ gives a typical 
Eden-like pattern of {\it B. subtilis} colonies. The mixture is autoclaved at 
121 $^\circ$C for 15 min, and 20 mL of the solution is poured into 
each sterilized plastic petri dish of 88 mm inner diameter. 
The thickness of the semisolid agar plates is about 3.2 mm. 
After solidification at room temperature for 60 min, 
the semisolid agar plates are dried at 50 $^\circ$C for 90 min.

3 $\mu$L of the bacterial suspension is inoculated at the 
center of each agar plate surface. 1 $\mu$L of the suspension 
includes about $10^2$ cells in the spore state. 
The agar plates are left at room temperature for about 60 min
to dry the bacterial suspension droplet. Thereafter, they are cultivated 
in a stage top incubator at 35 $^\circ$C (INULG2-OTOR-CV, Tokai Hit, Shizuoka),
which is attached to the stage of an optical microscope (IX71, Olympus, Tokyo). 
The spores at the inoculated center spot germinate about 2 h
after the inoculation. They repeatedly undergo cell multiplications without
splitting or movement, and form a long filament.
The filament of cells linked in tandem grows two-dimensionally 
on the agar plate surface like a self-elongating string. 
This growth process of a filament of cells is observed 
through an optical microscope with 
a 20x objective.

A digital camera (DP71, Olympus, Tokyo) is connected to
the optical microscope.
Microscopy snapshots are recorded every minute for about 6 h
using bio-imaging analysis software
(Lumina Vision, Mitani, Fukui and Tokyo)
from the time when spore germination occurred 
to the time when the entanglement of a filament started 
two-dimensional colony expansion.

%%%%%%%%%%%%%%%%%%%%%%%%%%%%%%%%%%%%%%%%%%%%%%%%%%%%%%%%%%
%%%  SEC3   %%%%%%%%%%%%%%%%%%%%%%%%%%%%%%%%%%%%%%%%%%%%%%
%%%%%%%%%%%%%%%%%%%%%%%%%%%%%%%%%%%%%%%%%%%%%%%%%%%%%%%%%%
\SSC{Crossover from One-Dimensional Structure
to Two-Dimensional Structure}
\label{sec:crossover}
%%%%%%%%%%%%%%%%%%%%%%%%%%%%%%%%%%%%%%%%%%%%%%%%%%%%%%%%%%

In this section, the time after the inoculated spot germinated 
is denoted by $T$ [min].
Let $L(T)$ [$\mu$m] be the total length of a filament of bacterial cells
observed at time $T$. If the specific growth rate is denoted by
$\mu$, it will show an exponential elongation,
\begin{equation}
L(T)=L(0) e^{\mu T}, \quad T \geq 0.
\label{eqn:elongation}
\end{equation}
First, we confirmed that the observed $L(T)$ data
up to $T=200$ min are well described by the single exponential
function Eq. (\ref{eqn:elongation}), and the specific growth rate was
evaluated to be
\begin{equation}
\mu=3.58 \times 10^{-2} \, \mbox{min$^{-1}$}.
\label{eqn:alpha_bar}
\end{equation}
This gives a cycle time (doubling time) of
\begin{equation}
\tau= \frac{\ln 2}{\mu}=19.4 \, \mbox{min}.
\label{eqn:tau}
\end{equation}

%%%%%%%%%%%%%%% Figure  %%%%%%%%%%%%%%%%%%%%%%%%%%%%%%%%%%%%%%%%%
\begin{figure}
\hskip 2.8cm
\includegraphics[width=0.7\linewidth]{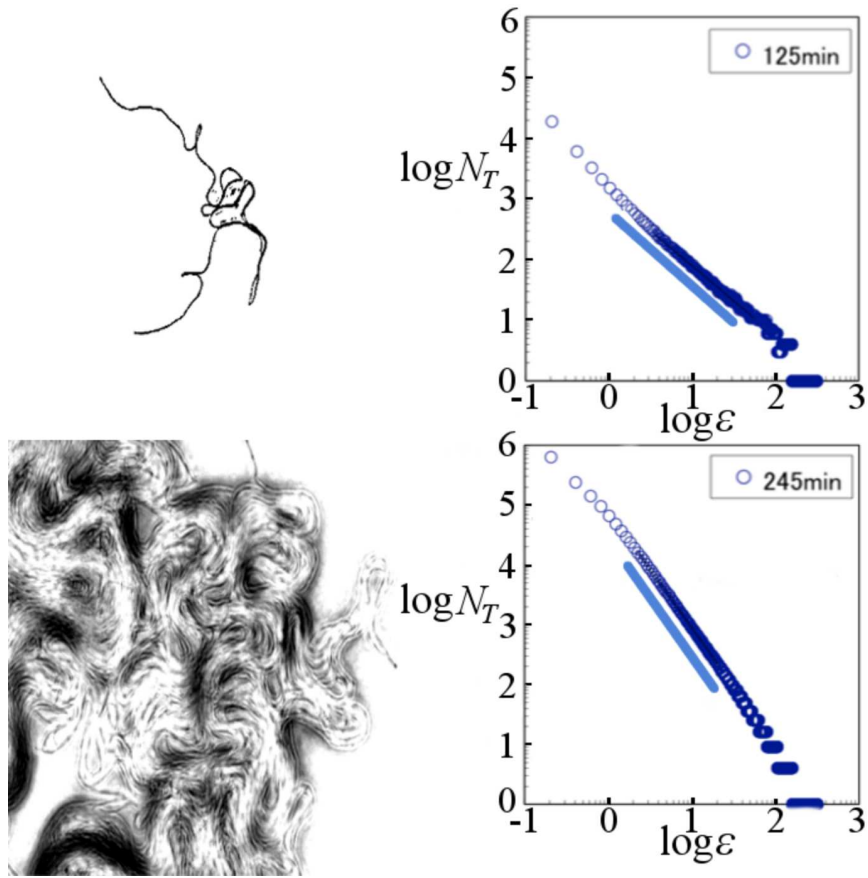}
\caption{
(Color online)
Upper left (lower left) picture:
snapshot of the cell filament at time
$T=125$ min ($T=245$ min).
The upper right (lower right) graph shows
the log-log plots of $N_T(\varepsilon)$ versus $\varepsilon$.
We find linear regions in the log-log plots,
and the fractal dimensions of the filament configurations
are evaluated as $D(125)=1.16$ and $D(245)=1.96$, respectively.
}
\label{fig:fractal1}
\end{figure}
%%%%%%%%%%%%%%%%%%%%%%%%%%%%%%%%%%%%%%%%%%%%%%%%%%%%%%%%%%

%%%%%%%%%%%%%%% Figure  %%%%%%%%%%%%%%%%%%%%%%%%%%%%%%%%%%%%%%%%%
\begin{figure}
\hskip 4.5cm
\includegraphics[width=0.4\linewidth]{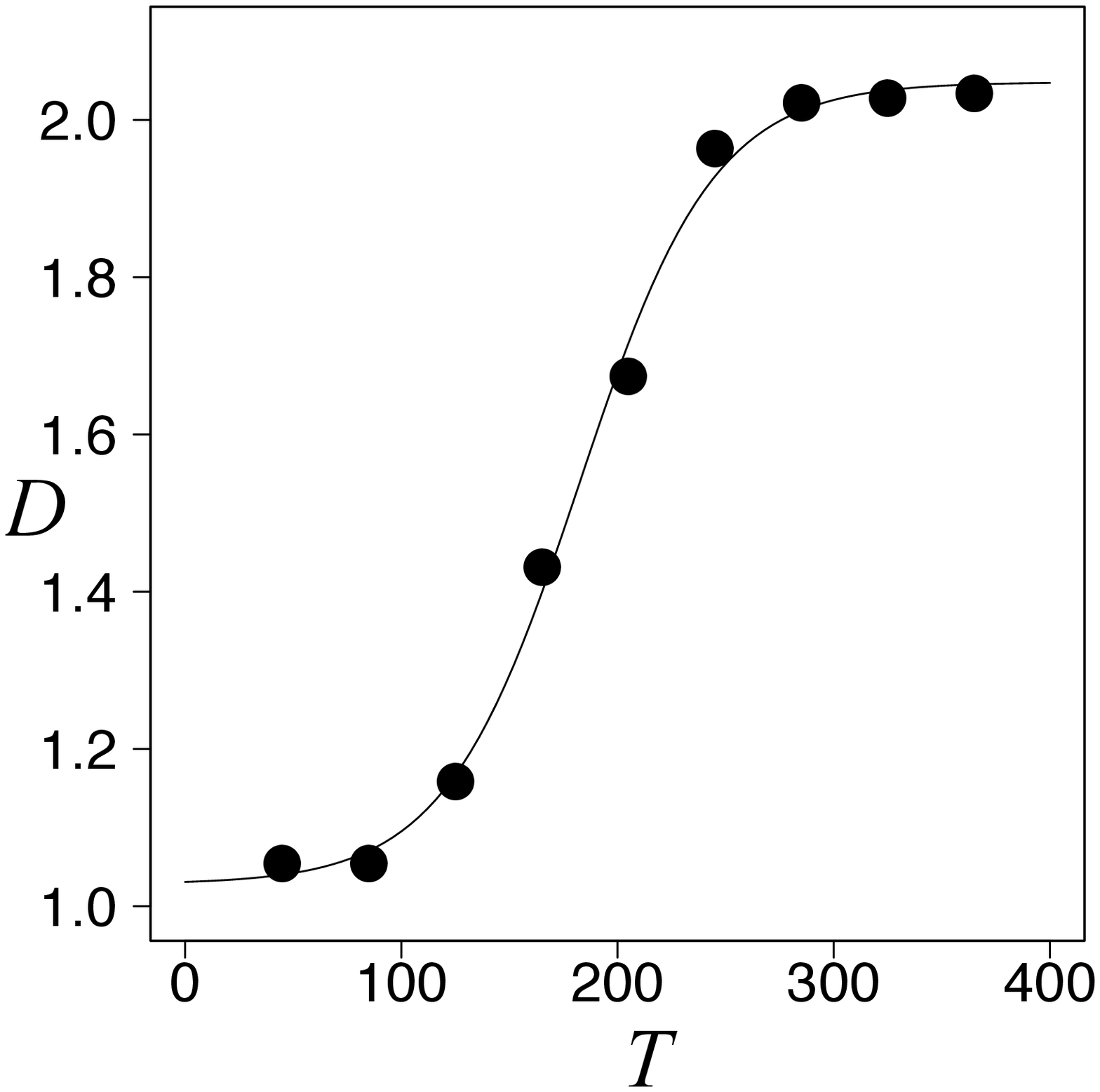}
\caption{
Time dependence of fractal dimension $D(T)$
of the filamentous cell configuration on the agar plate. 
The data are well described by
the sigmoid function given by Eq. (\ref{eqn:sigmoid}). 
}
\label{fig:fractal2}
\end{figure}
%%%%%%%%%%%%%%%%%%%%%%%%%%%%%%%%%%%%%%%%%%%%%%%%%%%%%%%%%%

The snapshots of filament configurations at
$T=125$ and 245 min are shown on the left-hand side in
Fig. \ref{fig:fractal1} with a resolution of $2040 \times1536$ pixels. 
They are analyzed by the box-counting method to evaluate the fractal 
dimensions of the filament configurations of bacterial cells on the agar plate. 
The procedure is as follows. 
At each time $T$, the snapshot picture is divided by
squares (two-dimensional boxes) of linear size $\varepsilon$, 
and then is counted the number $N_T(\varepsilon)$ of squares containing the pixels 
occupied by the filament of bacterial cell.
 If the configuration has a fractal structure, $N_T(\varepsilon)$ is scaled as
\begin{equation}
   N_T(\varepsilon) \sim \varepsilon^{-D(T)}
\label{eqn:fractal1}
\end{equation}
with the fractal dimension $D(T)$. 
We changed the value of $\varepsilon$ from 
1 to 1536 pixels (from $2.1 \times 10^{-1}$
to $3.2 \times 10^2$ $\mu$m in the real scale).
As shown in the log-log plots given on the right-hand side in Fig. \ref{fig:fractal1}, 
the data for $T=125$ min show a power law 
in the range of 3.5 $\mbox{$\mu$m} < \varepsilon < 5.6 \times 10^1$ \mbox{$\mu$m}
in the real scale,  and 
those for $T=245$ min show a power law
in the range of 2.5 $\mbox{$\mu$m} < \varepsilon < 2.1 \times 10^1$ \mbox{$\mu$m}
in the real scale.
The slopes $-1.16$ and $-1.96$ in these log-log plots in
Fig. \ref{fig:fractal1} provide the fractal dimensions 
$D(125)=1.16$ and $D(245)=1.96$, respectively. 
The result implies that the filament configurations have fractal structures. 
We evaluated the fractal dimensions of the filament 
at $T=45, 85, 165, 205, 285, 325$, and 365 min
and the results are plotted in Fig. \ref{fig:fractal2}.
We found that the data can be fitted to the following
sigmoid function:
\begin{equation}
D(T)=c_1 \tanh\bigl[\sigma(T-T_0)\bigr]+c_2
\label{eqn:sigmoid}
\end{equation}
with $T_0=1.82 \times$ $10^{2}$ min,
$\sigma=1.61 \times 10^{-2}$ \mbox{min$^{-1}$},
$c_1=5.1 \times 10^{-1}$, and $c_2=1.54$.
The time evolution of the multiple fission of bacterial cells
from the exponential phase to the stationary phase
is well described by the time dependence
of the fractal dimension of 
the filament configuration on the agar plate.

%%%%%%%%%%%%%%%%%%%%%%%%%%%%%%%%%%%%%%%%%%%%%%%%%%%%%%%%%%
%%%  SEC4   %%%%%%%%%%%%%%%%%%%%%%%%%%%%%%%%%%%%%%%%%%%%%%
%%%%%%%%%%%%%%%%%%%%%%%%%%%%%%%%%%%%%%%%%%%%%%%%%%%%%%%%%%
\SSC{Folding Processes and Experimental Measurements}
\label{sec:measurement}
%%%%%%%%%%%%%%%%%%%%%%%%%%%%%%%%%%%%%%%%%%%%%%%%%%%%%%%%%%

In this section, we denote the observation time by $t$ [min] instead of $T$ [min],
since we will set $t=0$ at the time when the first folding of a filament
occurs as explained below.

Let $L(t)$ [$\mu$m] be the total length of a filament of bacterial cells at time $t$.
Here, we write the specific growth rate as $\alpha$, 
and then $L(t)$ should obey the following differential equation:
\begin{equation}
\frac{d}{dt} L(t)=\alpha L(t).
\label{eqn:Lt2}
\end{equation}
On the semisolid agar plate,
the linear growth is unstable, and a filament of cells
starts folding after a short time.
It repeats the folding processes and the configuration of
the filament on the plate becomes complicated very rapidly.
Over time, the multifold structure becomes
dense  and its region spreads over the plates.
In this way, the configuration of a filament of cells
shows a crossover from the one-dimensional structure
to the two-dimensional structure, as shown in the previous section.

%%%%%%%%%%%%%%% Figure  %%%%%%%%%%%%%%%%%%%%%%%%%%%%%%%%%%%%%%%%%
\begin{figure}
\hskip 5cm
\includegraphics[width=0.4\linewidth]{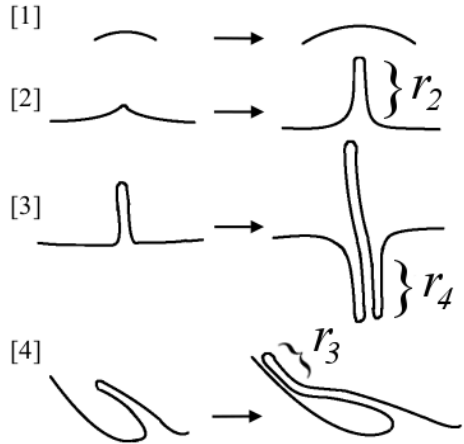}
\caption{
Illustrations of elementary processes found in the
self-elongating process of a filament of bacterial cells
in the early stage.
[1] Elongation of simple segment.
[2] Creation of a twofold segment by folding a simple segment.
Twice the length $r_2$ is added to the length $l_2$.
[3] Creation of a fourfold segment by elongation of a twofold segment.
Four times the length $r_4$ is added to the length $l_4$.
[4] Creation of a threefold segment by folding of a twofold segment
on a simple segment.
Three times the length $r_3$ is added to the length $l_3$.
}
\label{fig:processes}
\end{figure}
%%%%%%%%%%%%%%%%%%%%%%%%%%%%%%%%%%%%%%%%%%%%%%%%%%%%%%%%%%

In an interval of a filament, if bacterial cells are linked in tandem
and form a curved line without folding, 
the interval is called a simple segment.
At each time $t$, we consider the union of all simple segments
in a filament and call it the simple part.
We represent its length by $l_1(t)$ [$\mu$m].
As illustrated by [1] in Fig. \ref{fig:processes}, we see the following process:
$$
\mbox{Elementary Process [1] : elongation of a simple segment}, 
$$
and $l_1(t)$ rapidly increases with time $t$.

We set $t=0$ at the time when the first folding
of a filament is observed.
In this experiment, $t=T-60$ [min].
For $t >0$, the following process occurs:
\begin{eqnarray}
\mbox{Elementary Process [2]} &:& \mbox{creation of a twofold segment by folding of}
\nonumber\\
&& \mbox{a simple segment,}
\nonumber
\end{eqnarray}
as illustrated by [2] in Fig. \ref{fig:processes}.
The twofold part is defined by the union of all twofold segments
in the filament, whose length is denoted by $l_2(t)$ [$\mu$m].
By definition, $l_1(t)=L(t), l_2(t)=0$ for $t \leq 0$,
while $l_1(t)=L(t)-l_2(t), l_2(t)>0$
for $t > 0$.
Sooner or later, we will see threefold segments, fourfold segments,
and so forth.
We call the union of all $k$-fold segments the $k$-fold part, 
and write the length of the $k$-fold part
as $l_k(t)$ [$\mu$m], $k =2, 3, 4, \dots$. See the elementary processes 
[3] and [4] in Fig. \ref{fig:processes},
which create a fourfold segment and a threefold segment, respectively.
We will attempt to characterize the time evolution
of the filament configuration of cells using the set of lengths
$(l_1(t), l_2(t), l_3(t), \cdots)$ developing in time $t$.
Since $\sum_{k \geq 1} l_k(t)=L(t)$, 
the data are regarded as a time-dependent
`partition' of an exponentially growing length $L(t)$.

We found a critical time $t_*$ such that
when $0 \leq t \leq t_*$, the whole filament of cells
consists of only simple segments and twofold segments, 
while when $t > t_*$, 
we observe the appearance of 
threefold and fourfold segments in a filament of cells
and the configuration starts to become complicated.
In this experiment, we observed
\begin{equation}
t_*=45 \, \mbox{min}.
\label{eqn:t_*}
\end{equation}
It corresponds to the time 
$T_*=t_*+60 =105$ min
after the inoculated spot germinated.
Note that it gives the time when the fractal dimensions $D(T)$
of the filament starts to show its rapid increase
in Fig. \ref{fig:fractal2}. 

\begin{figure}
\hskip 2cm
\includegraphics[width=0.8\linewidth]{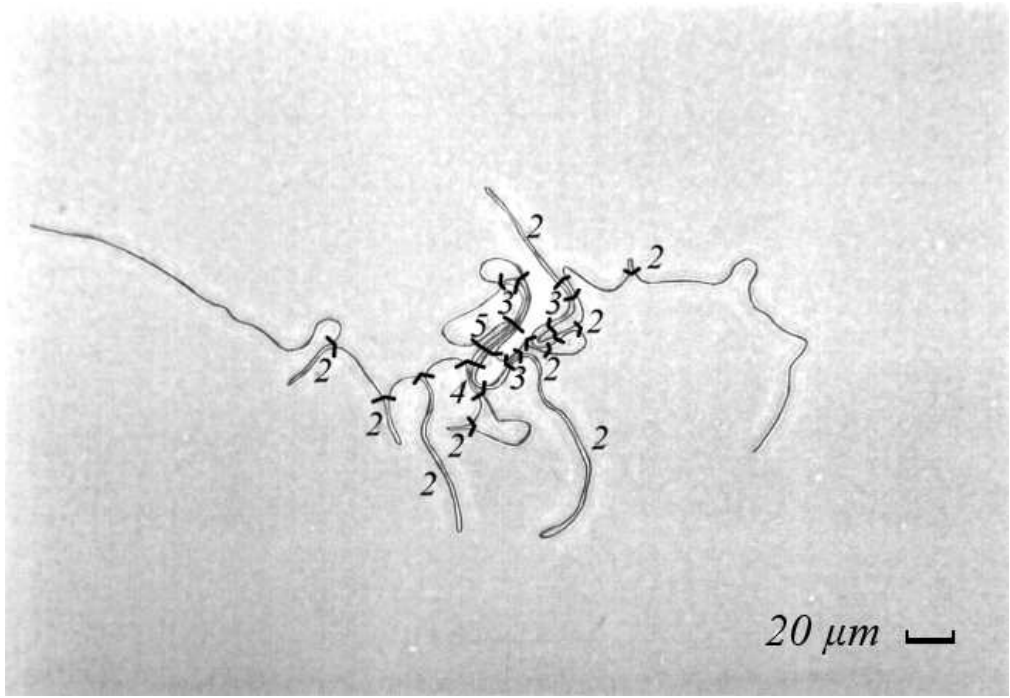}
\caption{Snapshot of a configuration of the filament of
bacterial cells at time $t=60$ min.
Each interval with a number $k$ represents
a $k$-fold segment of the filament, where $k=2, 3, 4$, and 5.
The intervals without a number are the simple segments.
The scale bar indicates 20 $\mu$m.
}
\label{fig:picture2}
\end{figure}

Figure \ref{fig:picture2} shows a snapshot 
of a configuration of bacterial cells at $t=60$ min.
We obtained an enlarged photocopy of the snapshot picture and
traced the filament segments by hand.
Each interval with a number $k$ represents
a $k$-fold segment of the filament, where $k=2,3,4$, and 5.
The intervals without a number are the simple segments.
We used an opisometer, which is an instrument
for measuring the length of arbitrary curved lines on a sheet.
If the length of a segment indexed $k$ is $r_k$, 
the length $l_k(t)$ of the $k$-fold part of the filament should be 
the sum of $k r_k$ over all $k$-fold segments.
We measured $(l_k(t))_{k \geq 1}$ 
up to time $t=90$ min.
The results are listed in Table \ref{tab:data},
where the lengths of parts with $k \geq 3$ are
summed and the values of $l_{3_+}(t)=\sum_{k \geq 3} l_{k}(t)$
are given.
In the experiment, we fixed the field of vision of the
optical microscope.
Just before $t=70$ min, a tip of the simple part
ran out of the field of vision, and after $t=70$ min,
twofold segments may have been created out of our field of vision.
Hence, the values of $L$ and $l_1$ at $t=70, 80$, and $90$ and those
of $l_2$ at $t=80$ and 90 are not given in Table \ref{tab:data}.

%%%%%%%%%%%%%%% Figure %%%%%%%%%%%%%%%%%%%%%%%%%%%%%%%%%%%%%%%%%
%%%%%%%%%%%%%%%%%%%%%%%%%%%%%%%%%%%%%%%%%%%%%%%%%%%%%%%%%%
%%%%%%%%%%%%% Table %%%%%%%%%%%%%%%%%%%%%%%%%%%%%%%%%%%%%%%%%
\begin{table}
\caption{Experimental data.}
\label{tab:data}
\begin{center}
\begin{tabular}{|r||r||r|r|r|}
\hline
$t$ [min] & $L$[$\mu$m] & $l_1$[$\mu$m] & $l_2$[$\mu$m] & $l_{3_+}$[$\mu$m] \\
\hline
0 &    151 &   151 &   0 &  0 \\
\hline
10 &   228 & 189 &  39 &  0 \\
\hline
20 &   315 & 247 &   68 &  0 \\
\hline
30 &   485 & 340 &   145 &  0 \\
\hline
40 &   733 & 442 & 291 &   0 \\
\hline
50 &   1070 & 620 & 362 &   90 \\
\hline
60 & 1590 & 810 & 584 &  197 \\
\hline
70 &  --- &   --- & 944 &  535 \\
\hline
80 &  --- &  --- & --- & 1260 \\
\hline
90 &  --- &  --- & --- & 2710 \\
\hline
\end{tabular}
\end{center}
\end{table}

%%%%%%%%%%%%%%%%%%%%%%%%%%%%%%%%%%%%%%%%%%%%%%%%%%%%%%%%%%
%%%  SEC5  %%%%%%%%%%%%%%%%%%%%%%%%%%%%%%%%%%%%%%%%%%%%%%
%%%%%%%%%%%%%%%%%%%%%%%%%%%%%%%%%%%%%%%%%%%%%%%%%%%%%%%%%%
\SSC{Analysis by Systems of Differential Equations}
\label{sec:analysis}
%%%%%%%%%%%%%%%%%%%%%%%%%%%%%%%%%%%%%%%%%%%%%%%%%%%%%%%%%
%%%%%%%%%%%%%%%%%%%%%%%%%%%%%%%%%%%%%%%%%%%%%%%%%%%%%%%%%
\subsection{Exponential growth of total length}
\label{sec:exponential}
%%%%%%%%%%%%%%%%%%%%%%%%%%%%%%%%%%%%%%%%%%%%%%%%%%%%%%%%
Equation (\ref{eqn:Lt2}) is solved as
\begin{equation}
L(t)=Ae^{\alpha t}.
\label{eqn:L2}
\end{equation}
Here, $A=L(0)$ is the total length of the filament 
at $t=0$ when the first folding occurs.

\begin{figure}
\hskip 4cm
\includegraphics[width=0.5\linewidth]{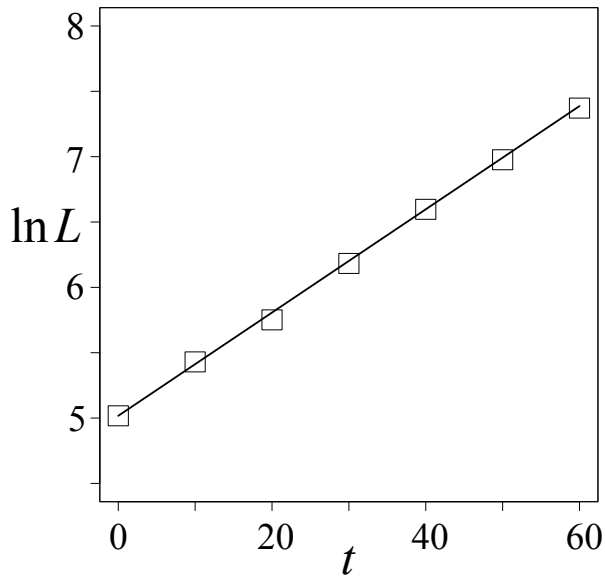}
\caption{Values of $\ln L(t)$ plotted for
$t=0,10, 20, \dots, 60$ min.
The linear fitting of Eq. (\ref{eqn:L2}) in this semilog plot
determines the values of
$\alpha=3.95 \times 10^{-2}$ \mbox{min$^{-1}$} and $A=1.50 \times 10^2$ $\mu$m.
}
\label{fig:exponential}
\end{figure}

As shown in Fig. \ref{fig:exponential}, the time dependence
of $L(t)$ is described by Eq. (\ref{eqn:L2}) very well 
for $0 \leq t \leq 60$ min, 
and we obtained the following values by
the semilog fitting of the data:
\begin{equation}
\alpha=3.95 \times 10^{-2} \, \mbox{min$^{-1}$}, \quad
A=1.50 \times 10^2 \, \mbox{$\mu$m}.
\label{eqn:fitting1}
\end{equation}
The evaluation of $\alpha$ is consistent with the evaluation
given by Eq. (\ref{eqn:alpha_bar}) for the specific growth rate
$\mu$ averaged over the longer time period 
$0 \leq T \leq 200$ min.

%%%%%%%%%%%%%%% Figure %%%%%%%%%%%%%%%%%%%%%%%%%%%%%%%%%%%%%%%%%
%%%%%%%%%%%%%%%%%%%%%%%%%%%%%%%%%%%%%%%%%%%%%%%%%%%%%%%%%%

%%%%%%%%%%%%%%%%%%%%%%%%%%%%%%%%%%%%%%%%%%%%%%%%%%%%%%
\subsection{Systems of differential equations and their solutions}
\label{sec:diff_eq}
%%%%%%%%%%%%%%%%%%%%%%%%%%%%%%%%%%%%%%%%%%%%%%%%%%%%%%%%%

In the time interval $0 \leq t \leq t_*$, 
only elementary processes [1] and [2] take place.
We assume that in the elongation process of a cell filament
with specific growth rate $\alpha$, 
the ratio of the frequency of elementary process [2]
to that of elementary process [1] is given by 
$\beta/(1-\beta)$ with a constant $0 < \beta < 1$.
Then, the time evolution of the lengths $l_1(t)$ and $l_2(t)$ 
will be described by
the following system of linear differential equations:
\begin{eqnarray}
\frac{d}{dt} l_1(t) &=& \alpha(1-\beta) l_1(t),
\nonumber\\
\frac{d}{dt} l_2(t) &=& \alpha l_2(t)+ \alpha \beta l_1(t),
\quad 0 \leq t \leq t_*.
\label{eqn:diff_eq1}
\end{eqnarray}
Note that the first term in the second equation in Eq. (\ref{eqn:diff_eq1})
describes the self-elongation process of the twofold part. 
Since we have set $t=0$ at the time when the first folding
occurs, Eq. (\ref{eqn:diff_eq1}) should be solved under
the conditions
\begin{equation}
l_1(0)=L(0)=A, \quad
l_2(0)=0.
\label{eqn:condition1}
\end{equation}
The solution is then given by
\begin{eqnarray}
l_1(t) &=& A e^{\alpha(1-\beta) t},
\nonumber\\
l_2(t) &=& A e^{\alpha t} (1-e^{-\alpha \beta t}),
\quad 0 \leq t \leq t_*.
\label{eqn:solutionA}
\end{eqnarray}

For $t > t_*$, we take into account the following elementary processes
in addition to processes [1] and [2]:
\begin{eqnarray}
\mbox{Elementary Process [3]} &:& \mbox{creation of a fourfold segment 
by elongation of} 
\nonumber\\
&& \mbox{a twofold segment}, 
\nonumber\\
\mbox{Elementary Process [4]} &:& \mbox{creation of a threefold segment
by folding of}
\nonumber\\
&& \mbox{a twofold segment on a simple segment}.
\nonumber
\end{eqnarray}
As illustrated by [3] in Fig. \ref{fig:processes},
a part of the elongating twofold segments becomes
a fourfold segment.
We assume that the ratio of the frequency of
fourfold segment creation to that of
simple elongation of the twofold part is given by 
$\gamma/(1-\gamma)$ with 
a constant $0 < \gamma < 1$.
As illustrated by [4] in Fig. \ref{fig:processes},
the creation of threefold segments can occur only if
a twofold segment touches a simple segment
and folds on it and if they merge into a threefold segment.
Thus, its frequency will be proportional to the product
of $l_1(t)$ and $l_2(t)$.
We assume that this process reduces the total
length of the simple part by
$\delta_1 l_1(t) l_2(t)$
and that of the twofold part
by $\delta_2 l_1(t) l_2(t)$
with transition rates per unit length
$\delta_1>0$ and $\delta_2 > 0$.
Then, if we set
$l_{3_+}(t)=\sum_{k \geq 3} l_k(t)=l_3(t)+l_4(t)+\cdots$,
we will obtain the following system of
nonlinear differential equations:
\begin{eqnarray}	
\frac{d}{dt}l_1(t) &=& \alpha(1-\beta)l_1(t)-\delta_1l_1(t)l_2(t), 
\nonumber\\
\frac{d}{dt}l_2(t) &=& \alpha(1-\gamma)l_2(t)+\alpha \beta l_1(t)-\delta_2l_1(t)l_2(t), 
\nonumber\\
\frac{d}{dt}l_{3_+}(t) &=& \alpha l_{3_+}(t)+\alpha \gamma l_2(t)+(\delta_1+\delta_2)l_1(t)l_2(t),
\quad t \geq t_*.
\label{eqn:diff_eq2}
\end{eqnarray}

We assume that the parameters $\delta_1$ and $\delta_2$
are sufficiently small and solve the system of
nonlinear differential equations given by Eq. (\ref{eqn:diff_eq2})
by perturbation.
This assumption will be verified by the data fitting as explained 
in Sect. \ref{sec:fittings}.

For $k=1,2$, and $3_+$, we expand $l_k(t)$
as power series of $\delta_1$ and $\delta_2$ as
\begin{eqnarray}
l_k(t) &=& \sum_{m_1=0}^{\infty} \sum_{m_2=0}^{\infty}
\delta_1^{m_1} \delta_2^{m_2} 
\tilde{l}_k^{(m_1, m_2)}(t)
\nonumber\\
&=& \sum_{n=0}^{\infty} 
\sum_{\substack{m_1 \geq 0, m_2 \geq 0, \cr m_1+m_2=n}}
\delta_1^{m_1} \delta_2^{m_2} \tilde{l}_k^{(m_1, m_2)}(t),
\label{eqn:expand}
\end{eqnarray}
where $\tilde{l}_k^{(m_1, m_2)}(t)$ are time-dependent coefficients
of the expansion.
The $N$th-order approximate solution, $N=0,1,2, \dots$, is
given by
\begin{equation}
l_k^{(N)}(t)=  \sum_{n=0}^{N} 
\sum_{\substack{m_1 \geq 0, m_2 \geq 0, \cr m_1+m_2=n}}
\delta_1^{m_1} \delta_2^{m_2} \tilde{l}_k^{(m_1, m_2)}(t),
\quad k=1,2,3_+.
\label{eqn:pth_appr}
\end{equation}
In the following, we calculate the 0th- and first-order approximate solutions:
\begin{eqnarray}
l_k^{(0)}(t) &=& \tilde{l}_k^{(0,0)}(t), 
\nonumber\\
l_k^{(1)}(t) &=& l_k^{(0)}(t)+\delta_1 \tilde{l}_k^{(1,0)}(t)
+ \delta_2 \tilde{l}_k^{(0,1)}(t),
\quad k=1,2,3_+, \quad t \geq t_*.
\label{eqn:0_1_appr}
\end{eqnarray}

The 0th-order approximate solution,
$\{l_k^{(0)}(t) : k=1,2,3_+\}$, solves
the system of linear differential equations
obtained from Eq. (\ref{eqn:diff_eq2}) by setting $\delta_1=\delta_2=0$.
In addition to the initial conditions corresponding to Eq. (\ref{eqn:condition1}),
\begin{equation}
l_1^{(0)}(0)=A, \quad l_2^{(0)}(0)=0,
\label{eqn:condition2}
\end{equation}
the definition of the critical time $t_*$ gives
\begin{equation}
l_{3_+}^{(0)}(t_*)=0.
\label{eqn:condition3}
\end{equation}
Under these conditions, we have the following:
\begin{eqnarray}
l_1^{(0)}(t) &=& A e^{\alpha(1-\beta) t},
\nonumber\\
l_2^{(0)}(t) &=& \frac{\beta A}{\beta-\gamma} e^{\alpha(1-\gamma) t}
(1-e^{-\alpha(\beta-\gamma)t} ),
\nonumber\\
l_{3_+}^{(0)}(t) &=&
\frac{A}{\beta-\gamma} e^{\alpha t}
\Big\{ -\beta(e^{-\alpha \gamma t} - e^{-\alpha \gamma t_*})
+\gamma(e^{-\alpha \beta t}-e^{-\alpha \beta t*}) \Big\},
\quad t \geq t_*.
\label{eqn:solution_0th}
\end{eqnarray}

To express the first-order approximate solution, 
we introduce the multiple integrals
\begin{eqnarray}
I^{(1)}(t; t_*, a_1)
&=& \int_{t_*}^t ds \, e^{-a_1 s} l_1^{(0)}(s) l_2^{(0)}(s),
\nonumber\\
I^{(2)}(t; t_*, a_2, a_1)
&=& \int_{t_*}^t ds \, e^{-a_2 s} I^{(1)}(s; t_*, a_1),
\nonumber\\
I^{(3)}(t; t_*, a_3, a_2, a_1)
&=& \int_{t_*}^t ds \, e^{-a_3 s} I^{(2)}(s; t_*, a_2, a_1),
\label{eqn:integrals}
\end{eqnarray}
where $a_i,$ $i=1,2$, and $3$ are constants.
By inserting $l_1^{(0)}(t)$ and $l_2^{(0)}(t)$ given in 
Eq. (\ref{eqn:solution_0th}), they are calculated as
\begin{eqnarray}
I^{(1)}(t; t_*, a_1)
&=& \frac{\beta A^2}{\beta-\gamma}
\left\{ \
\frac{e^{(p-a_1)t}-e^{(p-a_1)t_*}}{p-a_1}
-\frac{e^{(q-a_1)t}-e^{(q-a_1)t_*}}{q-a_1} \right\},
\nonumber\\
I^{(2)}(t; t_*, a_2, a_1)
&=& \frac{\beta A^2}{\beta-\gamma}
\left\{ 
\frac{e^{(p-a_1-a_2)t}-e^{(p-a_1-a_2)t_*}}{(p-a_1)(p-a_1-a_2)}
-\frac{e^{(q-a_1-a_2)t}-e^{(q-a_1-a_2)t_*}}{(q-a_1)(q-a_1-a_2)}
\right.
\nonumber\\
&& \quad \left. 
+\frac{e^{-a_2 t}-e^{-a_2 t_*}}{a_2} \left(
\frac{e^{(p-a_1)t_*}}{p-a_1}-\frac{e^{(q-a_1)t_*}}{q-a_1} \right)
\right\},
\nonumber\\
I^{(3)}(t; t_*, a_3, a_2, a_1)
&=& \frac{\beta A^2}{\beta-\gamma} \left[
\left\{ 
\frac{e^{(p-a_1-a_2-a_3)t}-e^{(p-a_1-a_2-a_3)t_*}}{(p-a_1)(p-a_1-a_2)(p-a_1-a_2-a_3)} \right. \right.
\nonumber\\
&& \quad
- \frac{e^{(q-a_1-a_2-a_3)t}-e^{(q-a_1-a_2-a_3)t_*}}{(q-a_1)(q-a_1-a_2)(q-a_1-a_2-a_3)} 
\nonumber\\
&& \quad
\left. -
\frac{e^{-(a_2+a_3)t}-e^{-(a_2+a_3) t_*}}{a_2(a_2+a_3)}
\left( \frac{e^{(p-a_1) t_*}}{p-a_1} - \frac{e^{(q-a_1)t_*}}{q-a_1} \right)
\right\}
\nonumber\\
&& \quad
+\frac{e^{-a_3 t}-e^{-a_3 t_*}}{a_3}
\left\{ \frac{e^{(p-a_1-a_2)t_*}}{(p-a_1)(p-a_1-a_2)}
-\frac{e^{(q-a_1-a_2)t_*}}{(q-a_1)(q-a_1-a_2)} \right.
\nonumber\\
&& \qquad
\left. \left.
+ \frac{e^{-a_2 t_*}}{a_2}
\left( \frac{e^{(p-a_1)t_*}}{p-a_1}-\frac{e^{(q-a_1)t_*}}{q-a_1} \right) 
\right\} \right],
\label{eqn:integrals2}
\end{eqnarray}
where
$p=\alpha(2-\beta-\gamma), q=2 \alpha(1-\beta)$.

Then, the first-order approximate solution is given by
\begin{eqnarray}
l_1^{(1)}(t) &=& l_1^{(0)}(t)
-\delta_1 e^{\alpha(1-\beta)t}
I^{(1)}(t; t_*, \alpha(1-\beta)),
\nonumber\\
l_2^{(1)}(t) &=& l_2^{(0)}(t)
- \delta_1 \alpha \beta e^{\alpha(1-\gamma) t}
I^{(2)}(t; t_*, -\alpha(\gamma-\beta), \alpha(1-\beta))
\nonumber\\
&& \quad \quad \, \, - \delta_2 e^{\alpha(1-\gamma) t} I^{(1)}(t; t_*, \alpha(1-\gamma)),
\nonumber\\
l_{3_+}^{(1)}(t) &=& l_{3_+}^{(0)}(t)
\nonumber\\
&+&
\delta_1 \Big\{ -\alpha^2 \beta \gamma e^{\alpha t}
I^{(3)}(t; t_*, \alpha \gamma, -\alpha(\gamma-\beta), \alpha(1-\beta)) 
+e^{\alpha t} I^{(1)}(t; t_*, \alpha) \Big\}
\nonumber\\
&+& 
\delta_2 \Big\{ -\alpha \gamma e^{\alpha t}
I^{(2)}(t; t_*, \alpha, \alpha(1-\gamma)) 
+e^{\alpha t} I^{(1)}(t; t_*, \alpha) \Big\},
\quad t \geq t_*.
\label{eqn:solution_1st}
\end{eqnarray}
Note that, owing to the initial condition
$l_1^{(1)}(0)=l_1^{(0)}(0)=A$,
$l_1^{(1)}(t)$ given in the first line of Eq. (\ref{eqn:solution_1st})
becomes independent of $\delta_2$.

%%%%%%%%%%%%%%%%%%%%%%%%%%%%%%%%%%%%%%%%%%%%%%%%%%%%%%
\subsection{Nonlinear fitting}
\label{sec:fittings}
%%%%%%%%%%%%%%%%%%%%%%%%%%%%%%%%%%%%%%%%%%%%%%%%%%%%%%%%%

%%%%%%%%%%%%%%% Figure %%%%%%%%%%%%%%%%%%%%%%%%%%%%%%%%%%%%%%%%%
\begin{figure}
\hskip 4cm
\includegraphics[width=0.5\linewidth]{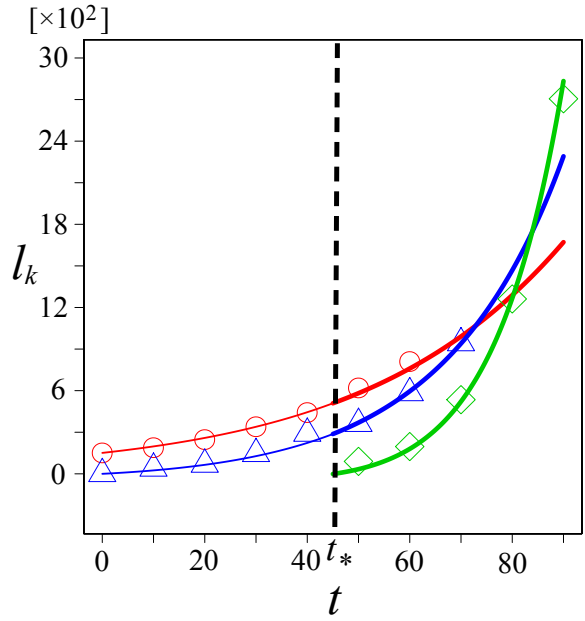}
\caption{
(Color online)
Data of $l_1(t)$, $l_2(t)$, and $l_{3_+}(t)$
plotted by $\bigcirc$, $\bigtriangleup$, and $\diamondsuit$,
respectively.
The thin curves represent the solution 
given by Eq. (\ref{eqn:solutionA}) for $0 \leq t \leq t_*$
with the parameters given by Eqs. (\ref{eqn:fitting1}) and (\ref{eqn:beta}).
The solution given by Eq. (\ref{eqn:solution_1st}) for $t \geq t_*$
is represented by the thick curves, where
the parameters are given by Eq. (\ref{eqn:fitting2}).
The fitting is excellent.
}
\label{fig:fitting1}
\end{figure}
%%%%%%%%%%%%%%%%%%%%%%%%%%%%%%%%%%%%%%%%%%%%%%%%%%%%%%%%%%
%%%%%%%%%%%%%%% Figure %%%%%%%%%%%%%%%%%%%%%%%%%%%%%%%%%%%%%%%%%
\begin{figure}
\hskip 4cm
\includegraphics[width=0.5\linewidth]{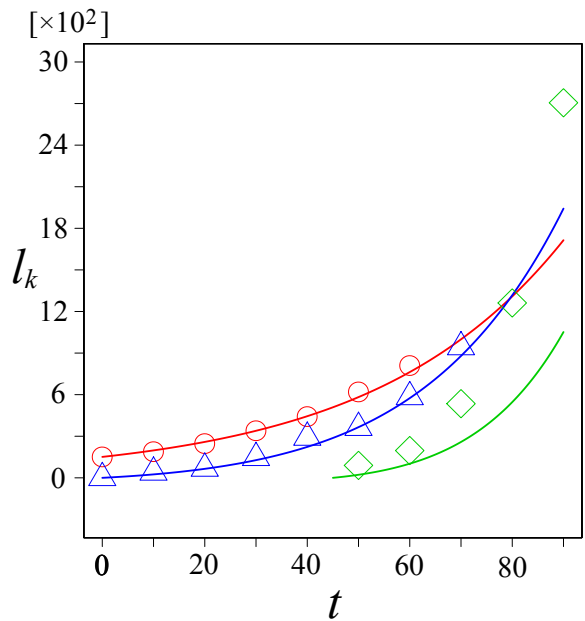}
\caption{
(Color online)
Data of $l_1(t)$, $l_2(t)$, and $l_{3_+}(t)$
plotted by $\bigcirc$, $\bigtriangleup$, and $\diamondsuit$,
respectively.
The curves show the 0th-order solution given by Eq. (\ref{eqn:solution_0th}),
which ignores the nonlinear terms in Eq. (\ref{eqn:diff_eq2}). 
We failed to fit $l_{3_+}(t)$.
}
\label{fig:fitting2}
\end{figure}
%%%%%%%%%%%%%%%%%%%%%%%%%%%%%%%%%%%%%%%%%%%%%%%%%%%%%%%%%%

First, we used five pairs of data $(l_1(t), l_2(t))$ of
$t=0,10, \dots, 40 < t_*=45$ in Table \ref{tab:data}.
Note that the parameters $\alpha$ and $A$ have already been
determined using Eq. (5.2).
By fitting to the solution given by Eq. (\ref{eqn:solutionA})
of the system given by Eq. (\ref{eqn:diff_eq1}) of linear differential equations 
for $(l_1(t), l_2(t)), 0 \leq t \leq t_*$, we obtained the value of parameter
\begin{equation}
\beta=0.317.
\label{eqn:beta}
\end{equation}

Next, we used the data $l_k(t)$, $k=1,2$, and $3_+$ for
$t=50, 60, \dots, 90 > t_*=45$ in Table \ref{tab:data}
and performed their nonlinear fitting
to the first-order approximate solution given by Eq. (\ref{eqn:solution_1st}) of the
system of nonlinear differential equations 
given by Eq. (\ref{eqn:diff_eq2}).
Here, $A, \alpha$, and $\beta$ are fixed to be the values
given by Eqs. (\ref{eqn:fitting1}) and (\ref{eqn:beta}),
and $\gamma, \delta_1$, and $\delta_2$ are chosen as
fitting parameters.
They are evaluated as 
\begin{eqnarray}
&& \gamma = 0.313, 
\nonumber\\
&& \delta_1=6.24 \times 10^{-7} \, \mbox{min$^{-1}\cdot\mu$m$^{-1}$}, \quad
\delta_2=3.75 \times 10^{-8} \, \mbox{min$^{-1}\cdot\mu$m$^{-1}$}.
\label{eqn:fitting2}
\end{eqnarray}
The fitting is excellent as shown by Fig. \ref{fig:fitting1}.
The evaluation given by Eq. (\ref{eqn:fitting2}) is
consistent with the assumption
$|\delta_i| \ll 1, i=1,2$, 
on which we solved the system of nonlinear differential
equations given by Eq. (\ref{eqn:diff_eq2}) by perturbation
in Sect. \ref{sec:diff_eq}.

The nonlinearity is very small but necessary in the fitting.
To demonstrate it, we show the 0th-order approximate solution
given by Eq. (\ref{eqn:solution_0th}) as curves in
Fig. \ref{fig:fitting2}.
Here, we used the same values of $A$, $\alpha$, and $\beta$
as in Fig. \ref{fig:fitting1}, but we put $\delta_1=\delta_2=0$.
Figure \ref{fig:fitting2}
shows that if we ignore the nonlinear terms
in Eq. (\ref{eqn:diff_eq2}), we fail to fit
the data of $l_{3_+}(t), t \geq t_*$.

%%%  SEC6   %%%%%%%%%%%%%%%%%%%%%%%%%%%%%%%%%%%%%%%%%%%%%%
%%%%%%%%%%%%%%%%%%%%%%%%%%%%%%%%%%%%%%%%%%%%%%%%%%%%%%%%%%
\SSC{Concluding Remarks}
\label{sec:remarks}
%%%%%%%%%%%%%%%%%%%%%%%%%%%%%%%%%%%%%%%%%%%%%%%%%%%%%%%%%%

In this paper, we have reported that
the growth process of cells of {\it B. subtilis}
under hard-agar and nutrient-rich conditions
allows the realization of the dynamics of
a self-elongating filament with sequential folding on a plane.
Such a multiple-fission process without cell separation
is commonly observed in the early stage of the growth process
even if the agar concentration is changed,
while the structure and motion of 
an entangled filament of cells in the later stage 
depend on environmental conditions
\cite{Men76,MSL97,WIMM97,WKMM10,KIT96}.
Takeuchi {\it et al.} reported a study on
filamentous cells of {\it E.coli} \cite{TDWW05}.
Environmental conditions to realize
such self-elongation of cell filaments
should be clarified by a systematic study of the early
stage of bacterial growth processes.
The classification of the morphology and dynamics
of a long filament of cells depending on
environmental conditions, time periods,
spatial and geometrical restrictions, and so forth
will be an interesting future problem.

Here, we have focused on the simplest situation
wherein a filament of cells simply repeats
folding processes as it elongates and
isotropically spreads over a two-dimensional plate.
Note that Mendelson and coworkers
have very extensively studied filamentous 
cell growth in the situation wherein
supercoiling processes create helical 
macrofibers and their chiral self-propulsion
motion is observed 
\cite{Men76,Men78, MTKL95,MSL97,Men99,
MSWG00,MST01,MMT02,MSRCT03}.
Even in our simple case, it seems to be highly
nontrivial to provide a proper description of
the filament configuration, which rapidly becomes
complicated as it elongates with sequential folding
embedded in a plane.
In this work, we have proposed describing
the global development by the time-dependent
fractal dimension $D(T)$ and the local folding processes
by the time evolution of partitions
$(l_k(t))_{k \geq 1}$ of the exponentially
growing total length $L(t)$ of the filament of
cells, where $k=1$ for the simple part
and $k \geq 2$ for the $k$-fold parts.

The analysis discussed in Sect. \ref{sec:analysis}
could be regarded as a mean-field-type approximation
in the following sense.
Let us consider a magnetic spin system on a lattice.
In the mean-field theory, to describe a phase transition,
we consider only the magnetization as an order parameter,
which is obtained by averaging over spin configurations.
The magnetization per spin $m(T,H)$ at temperature $T$ 
in an external magnetic field $H$ is calculated by approximating
the correlated many-spin system by 
a single-spin system in a mean field generated by the surrounding spins,
which is assumed to be proportional to $m(T, H)$.
The proportionality coefficient can be called 
the effective coupling constant $J_{\rm eff}$.
In this way, we obtain the self-consistency equation for $m(T, H)$
with the parameter $J_{\rm eff}$ in addition to the
external parameters $T$ and $H$.
If we want to compare the experimental data of
a magnetization process of some material with the mean-field theory,
the parameter $J_{\rm eff}$ should be evaluated by
some additional experimental observation.
In this analysis of the filament configuration with folding,
we summed $l_k(t)$ over $k \geq 3$
to define $l_{3_+}(t)$.
By this reduction of variables from the series $(l_k(t))_{k \geq 1}$
to the triplet $(l_1(t), l_2(t), l_{3_+}(t))$,
we obtained the finite systems of
coupled differential equations given by Eq. (\ref{eqn:diff_eq1}) 
for $0 \leq t \leq t_*$ and by Eq. (\ref{eqn:diff_eq2}) for $t \geq t_*$.
They involve the parameters
$\alpha, \beta, \gamma, \delta_1$, and $\delta_2$.
We have evaluated these parameters as well as $t_*$
by experimental observations.

As shown by Eqs. (5.16) and (5.17), the evaluated $\beta$ and $\gamma$
have almost the same value. We can verify that the quantities given by
Eqs. (5.12) and (5.14) have finite values in the limit $\gamma \to \beta$,
and then our first-order approximate solution given by Eq. (5.15) is
also valid in the case where $\beta=\gamma$. We should note that
if $\beta=1/3$ ($\gamma=1/3$), the ratio of the frequency
of twofold-segment (fourfold-segment) creation to that
of simple elongation of the single (twofold) part
is given by $\beta/(1-\beta)=1/2$ ($\gamma/(1-\gamma)=1/2$).
We will continue our study to answer the question whether this
result, $\beta \simeq \gamma \simeq 1/3$, is universal.

To improve the description,
we need to carry out further studies on growing elastic filaments.
Theoretical investigations can be found in the literature 
\cite{PHBC04,WGP04,GN06,MLG13}.
We hope that this experimental evaluations
of parameters controlling the folding processes,
which are described as
$l_1 \to l_2$, $l_2 \to l_4$, and
$l_1+l_2 \to l_3$ 
using our variables, 
will be useful for testing the validity of a possible theoretical consideration
in the future.

\vskip 0.5cm
%%%%%%%%%%%%%%%%%%%%%%%%%%%%%%%%%%%%%%%%%%%%%%%%%%%%%%
\noindent{\bf Acknowledgments} 
%%%%%%%%%%%%%%%%%%%%%%%%%%%%%%%%%%%%%%%%%%%%%%%%%%%%%%
JW was supported by a Chuo University Grant for Special Research and 
by a Grant-in-Aid for Exploratory Research (No. 15K13537)
from Japan Society for the Promotion of Science.
MK was supported in part by
a Grant-in-Aid for Scientific Research (C)
(No. 26400405) from Japan Society for
the Promotion of Science.
%%%%%%%%%%%%%%%%%%%%%%%%%%%%%%%%%%%%%%%%%%%%%%%%%%%%%%%%%%%%
%%%%%%%%% References %%%%%%%%%%%%%%%%%%%%%%%%%%%%%%%%%%%%%%%
%%%%%%%%%%%%%%%%%%%%%%%%%%%%%%%%%%%%%%%%%%%%%%%%%%%%%%%%%%%%

%%%%%%%%%%%%%%%%%%%%%%%%%%%%%%%%%%%%%%%%%%%%%%%%

\end{document}